\documentstyle[12pt]{article} \begin{document}
\def\<{\langle}
\def\>{\rangle}
\newcommand{\EQ}{\begin{equation}}
\newcommand{\EN}{\end{equation}}
\newcommand{\EQA}{\begin{eqnarray}}
\newcommand{\EQN}{\end{eqnarray}}
\newcommand{\EQAN}{\begin{eqnarray*}}
\newcommand{\EQNN}{\end{eqnarray*}}
\newcommand{\e}{{\rm e}}
\newcommand{\Sp}{{\rm Sp}}
\newcommand{\Tr}{{\rm Tr}}
\newcommand{\p}{\partial}
\newcommand{\h}{{1\over 2}}
\newcommand{\tright}{$\triangleright \quad$}
\newcommand{\tleft}{$\triangleleft \quad$}
\newcommand{\IR}{\relax{\rm I\kern-.18em R}}
\renewcommand{\theequation}{\arabic{equation}}


\begin{titlepage}
\begin{flushright}
  hep-th/9906248\\
YITP-99-40\\
EFI-99-30\\
USC-99/HEP-M4\\
UT-KOMABA/99-10\\
\end{flushright}
\begin{center}
{\bf ON THE QUANTIZATION OF NAMBU BRACKETS } \\
\vspace{1cm}
{\bf Hidetoshi Awata$^{a}$\footnote{e-mail: awata@yukawa.kyoto-u.ac.jp}
, Miao Li$^{b}$\footnote{e-mail: mli@theory.uchicago.edu} \\
\vspace{.5cm}
and\\
\vspace{.5cm}
 Djordje Minic$^{c}$\footnote{e-mail: minic@physics.usc.edu},
 Tamiaki Yoneya$^{d}$} \footnote{e-mail: tam@hep1.c.u-tokyo.ac.jp}  \\
\vspace{.5cm}
$^{a}$Yukawa Institute for Theoretical Physics, Kyoto
University, 606 Kyoto, Japan\\
\vspace{.1cm}
$^{b}$Enrico Fermi Institute, University of Chicago,
5640 Ellis Avenue, Chicago, IL 60637, USA \\
\vspace{.1cm}
$^{c}$Department of
Physics and Astronomy, University of Southern California, Los Angeles,
CA 90089-0484, USA \\
\vspace{.1cm}
$^{d}$Institute of Physics, University of Tokyo, Komaba,
Meguro-ku, 153 Tokyo, Japan \\
\vspace{.5cm}
{\bf Abstract}
\end{center}

We present several non-trivial examples of the three-dimensional quantum
Nambu bracket which
involve square matrices or three-index objects. Our examples satisfy
two fundamental properties of the classical Nambu bracket:
they are skew-symmetric and they obey the Fundamental Identity.
We contrast our approach
to the existing literature on the quantum deformations of Nambu mechanics.
We also discuss possible applications of our results in M-theory.

\end{titlepage}


\section{Introduction}


In 1973 Nambu \cite{nambu} proposed a generalization of Hamiltonian
mechanics and
statistical mechanics involving odd-dimensional ``phase-spaces".
In particular, in the case of a three-dimensional ``phase-space" labeled by
$x,y,z$ Nambu proposed the following generalized Hamilton evolution
equation
\EQ
{dF \over dt} = {\partial{(H_1,H_2,F)} \over \partial{(x,y,z)} }
\EN
where the right hand side (R.H.S.) denotes the three-dimensional Jacobian
of
$H_1(x,y,z)$, $H_2(x,y,z)$, $F(x,y,z)$ with respect to $x,y,z$ and defines  
the
classical Nambu bracket. Nambu's proposal was motivated by the general
validity of the Liouville theorem, as it is apparent from the form of
the evolution equation (1).

It was then noticed that various physical systems described by this
formalism
in $n$ ``phase-space" dimensions can
be realized as singular Hamiltonian systems in $2n$ ``phase-space"
dimensions \cite{flato1}.
The original proposal of Nambu has also been given an elegant geometric
formulation
by Takhtajan \cite{fi}.

The quantization problem of the classical Nambu bracket turned out to
be very difficult. The only existing approach uses an unusual
version of deformation
quantization called Zariski quantization \cite{flato}, \cite{flato2}.
However, an explicit realization of the quantum Nambu bracket in terms
of
matrices, as posed in the original paper by Nambu, still seems to be
lacking.

It is one of the aims of this article to
present several non-trivial examples of the quantum
version of the Nambu bracket in terms of ordinary matrices and also in
terms of three-index objects (cubic matrices).
Our main motivation for this investigation comes from an expectation that
the
quantum Nambu bracket might turn out to be a useful technical tool in
M-theory.

The article is organized as follows:
first, in section 2,  we briefly review
the connection between volume preserving
diffeomorphisms and the classical Nambu bracket,
and point out some novel structural
features of the latter.
Then in section 3, we give an explicit matrix realization of
the three-dimensional quantum Nambu bracket which
satisfies two crucial properties of its classical counterpart:
skew-symmetry and the Fundamental Identity. We contrast our results to the
existing examples obtained in the formalism of Zariski quantization
\cite{flato2}.
In section 4,
we give an explicit realization of the three-dimensional quantum Nambu
bracket in terms of three-index objects, which we call cubic matrices.
In section 5,  we present the generalization of our approach
to the $n$-dimensional quantum Nambu bracket.
Finally in section 6,  we discuss possible
applications of our results in M-theory.


\section{Volume Preserving Diffeomorphisms and
the Classical Nambu Bracket}


Consider a three-dimensional space parametrized by
$\{x^i\}$.
The three-dimensional volume preserving diffeomorphisms (VPD) on this space
are described by a differentiable map
\EQ
x^i \rightarrow y^i (x)
\EN
such that
\EQ
\{y^1, y^2, y^3 \} =1
\EN
where, by definition
\EQ
\{A, B, C\} \equiv \epsilon^{ijk}\partial_i A \partial_j B \partial_k C
\label{npbracket}
\EN
is the Nambu-Poisson (NP) bracket, or Nambu bracket, or
Nambu triple bracket, which satisfies \cite{fi}, \cite{flato},
\cite{flato2}, \cite{filip}
\begin{enumerate}
\item Skew-symmetry
\EQ
\{A_1, A_2,A_3\}=(-1)^{\epsilon(p)} \{ A_{p(1)}, A_{p(2)},A_{p(3)} \},
\label{skewsymmetry}
\EN
where $p(i)$ is the permutation of indices and
$\epsilon(p)$ is the parity of the permutation,
\item Derivation
\EQ
\{A_1A_2, A_3, A_4\} =A_1\{A_2, A_3,A_4\} + \{A_1, A_3,A_4\}A_2 ,
\EN
\item Fundamental Identity (FI-1) \cite{fi}, \cite{filip}
\EQA
\{\{A_1, A_2, A_3\},A_4, A_5 \} +\{A_3, \{A_1, A_2,A_4\},A_5\}
\nonumber \\
+\{A_3, A_4, \{A_1, A_2, A_5\}\} =
\{A_1,A_2,\{A_3, A_4, A_5\}\}.
\EQN
\end{enumerate}

The three-dimensional VPD involves two independent functions.
Let these functions be denoted by $f$ and $g$.
The infinitesimal three-dimensional VPD generator is then given as
\EQA
D{(f, g)} &\equiv& \epsilon^{ijk}\partial_i f \partial_j g \partial_k
\\
           &\equiv&  D^k(f, g) \partial_k .
\label{3dapdgenerator}
\EQN
The volume-preserving property is nothing but the identity
\EQ
\partial_i D^i(f,g) =\partial_k (\epsilon^{ijk}\partial_i f \partial_j
g )=0 .
\EN
Given an arbitrary scalar function $X(x^i)$, the three-dimensional VPD
acts as
\EQ
D{(f,g)}X = \{f, g, X\} .
\EN
Apart from the issue of global definition of the functions $f$ and $g$,
we can represent an arbitrary infinitesimal volume-preserving
diffeomorphism
in this form.

On the other hand, if the base three-dimensional space $\{x^i\}$
is mapped into a target space of dimension $d+1$ whose
coordinates are $X^{\alpha} \, \, (\alpha =0,1,2, \ldots, d)$,
the induced infinitesimal volume element is
\EQ
d\sigma \equiv \sqrt{
\{X^{\alpha}, X^{\beta}, X^{\gamma}\}^2
}dx^1dx^2dx^3,
\EN
provided the target space is a flat Euclidean space.
The volume element is of course invariant under the general
three-dimensional
diffeomorphisms.

The triple product $\{X^{\alpha}, X^{\beta}, X^{\gamma}\}$ is also
"invariant" under the VPD. Or more precisely, it transforms as a scalar.
Namely,
\EQ
\{Y^{\alpha}, Y^{\beta}, Y^{\gamma}\} -
\{X^{\alpha}, X^{\beta}, X^{\gamma}\} =\epsilon D(f,g)\{X^{\alpha},
X^{\beta}, X^{\gamma}\} +O(\epsilon^2)
\EN
for
\EQ
Y=X+ \epsilon D(f,g)X .
\EN
This is due to the the Fundamental Identity FI-1 which shows that the
operator $D{(f,g)}$ acts as a derivation within the
NP bracket. For fixed $f$ and $g$, we can define a finite transformation
by \EQ
X(t) \equiv \exp (tD(f,g)) \, \rightarrow X =
\sum_{n=0}^{\infty}
{t^n\over n!}\{f, g, \{f,g, \{ \ldots , \{f,g,\{f,g, X\}\}\ldots,\}\}\}
\EN
which satisfies the Nambu ``equation of motion" \cite{nambu}
\EQ
{d\over dt}X(t)=\{f,g, X(t)\}.
\EN
The Nambu-Poisson structure is preserved under this evolution equation.

Let us now consider the group property of the three-dimensional VPD.
First we derive, using the FI-1
\EQA
[D{(f_1, g_1)}, D{(f_2, g_2})] X
\equiv  \{f_1, g_1, \{f_2, g_2, X\}\} -\{f_2, g_2, \{f_1, g_1, X\}\}
\nonumber\\
= \{\{f_1, g_1,f_2\}, g_2, X\} +\{f_2, \{f_1, g_1,g_2\}, X\} .
\label{2ndline}
\EQN
The second  line, however, does not have manifest antisymmetry under
the interchange $1\leftrightarrow 2$. To understand what this means,
we note that
the Nambu bracket defined by (\ref{npbracket}) satisfies
the following identities in addition to the FI-1 as
pointed out by Hoppe \cite{hoppe}
\begin{enumerate}
\item FI-2
\EQ
\{A_{[1}, A_2,\{A_3, A_{4]}, B\}\} =0
\EN
\item FI-3
\EQA
\{\{A_1, A_2, A_3\}, A_4, B\} &-&\{\{A_2, A_3, A_4\}, A_1, B\}
\nonumber \\
+\{\{A_3, A_4, A_1\}, A_2, B\} &-&\{\{A_4, A_1, A_2\}, A_3,B\}=0 .
\EQN
\end{enumerate}
Here we changed the order of the elements, using skew-symmetry,
from Hoppe's original form.
It is claimed in \cite{hoppe} that two of  the identities FI-1,
FI-2, FI-3 are independent.
We wish to show that FI-2 and FI-3 can be derived
from FI-1 and skew-symmetry.

The identity FI-2 and skew-symmetry
property enable us to rewrite (\ref{2ndline}) as
\EQA
&\{\{f_1, g_1,f_2\}, g_2, X\} +\{f_2, \{f_1, g_1,g_2\}, X\} &=
-\{\{f_2, g_2,f_1\}, g_1, X\}
\nonumber\\
& -\{f_1, \{f_2, g_2,g_1\}, X\}.&
\label{antisymmetry}
\EQN
Thus we recover the required antisymmetry under $1\leftrightarrow 2$.

However, in our case, the left hand side (L.H.S.) of the first line in
(\ref{2ndline}) is
antisymmetric by definition, and therefore (\ref{antisymmetry})
{\it must} hold identically.  Thus we conclude that the
identity (\ref{antisymmetry})  is
actually a consequence of the original fundamental identity FI-1,
contrary to Hoppe's statement \cite{hoppe}.

Note that in going from the first line to the second line in
(\ref{2ndline}), the explicit form
of the Nambu bracket need not be used, nor the property of skew-symmetry.
If the property of skew-symmetry with respect to the first two entries is
assumed,
(\ref{antisymmetry}) is equivalent to the FI-3.
Hence, the FI-2 is also a consequence of the FI-1.

The algebra of $D(f,g)$ now takes the form
\EQA
[D(f_1, g_1), D(f_2, g_2)] &=&{1\over 2}\Bigl(
D(\{\{f_1, g_1\},g_1\}, g_2)+D(f_2, \{f_1, g_1,g_2\})
 \nonumber \\
&-&D(\{f_2, g_2,f_1\}, g_1)-D(f_1, \{f_2, g_2,g_1\})\Bigr).
\label{4thline}
\EQN
One important lesson  is that the skew-symmetry
property
(\ref{skewsymmetry}) is not necessary for the
group structure of the transformations generated by
$D(f,g)$. What is then the role of skew-symmetry
from the view point of symmetry? One
obvious fact is that
the property of skew-symmetry with respect to the first and the second
entries means that only the
``independent" part of the two parameter functions $f$ and $g$
contributes to the transformation, in the sense that
\EQ
D(f,g)=D(f+cg, g)=D(f, g+cf)
\EN
for arbitrary constant $c$.

In the case of the usual Poisson structure,
the algebra of two-dimensional area preserving diffeomorphisms is given by
\EQ
[D(f_1), D(f_2)] =D(f_3)
\EN
where
\EQ
f_3= \{f_1, f_2\}
\EN
\EQ
D(f)X =\{f, X\}
\EN
Formally, the three-dimensional VPD algebra should also  be
expressible in the form
\EQ
[D(f_1, g_1), D(f_2, g_2)] =D(f_3, g_3)
\EN
Is it indeed possible to express $(f_3,g_3)$ in terms of the
Nambu bracket? It turns out that the three-dimensional analogue
of the commutator algebra
\EQ
D(A_{[1}) D(A_{2]}) =D(\{ A_1, A_2 \})
\EN
can be written using the quantum triple Nambu commutator \cite{nambu}
\EQ
[A,B,C]_N \equiv ABC-ACB+BCA-BAC+CAB-CBA
\label{nambutriple}
\EN
as follows
\EQ
D(A_{[1}, A_2) D(A_{3]_N},B) = 2D(\{ A_1, A_2, A_3 \},B),
\EN
or equivalently
\EQ
D(B_{[1}, B_2) D(A_{[1},A_2) D(A_{3]_N},B_{3]_N})=
4D(\{ A_1, A_2, A_3 \},\{ B_1, B_2, B_3 \}).
\EN
Both relations are equivalent to the Fundamental Identity
\EQA
&\{ A_1, A_2, \{A_3,B,C \}
 \} + \{ A_2, A_3, \{A_1,B,C \}
 \} + \{ A_3, A_1, \{A_2,B,C \}
 \} &
\nonumber\\
&= \{ \{A_1, A_2, A_3 \},B,C \}&.
\EQN
This result suggests that there is a new kind of
symmetry based on a new composition law
whose infinitesimal algebra is given by the
triple commutator (\ref{nambutriple}).
As far as we know this type of symmetry
has not been observed previously in the literature. It is tempting to
conjecture that this symmetry is  related
to the gauge transformations that are not of Yang-Mills type as in
\cite{volume}.


\section{Square Matrices and the Quantum Nambu Brackets}


The problem of discretization of p-dimensional volume preserving
diffeomorphisms is related to the issue of
quantization of the p-dimensional Nambu bracket.
Apparently there exists in the literature only one solution of the
quantization
problem of Nambu brackets which is based on a non-standard
deformation quantization called Zariski quantization \cite{flato}.
The quantum Nambu bracket constructed in \cite{flato} is
skew-symmetric and obeys both the derivation property and the
Fundamental Identity.
There also exists an example of a quantum Nambu bracket constructed
via the same deformation quantization procedure, which is skew-symmetric
and obeys the Fundamental Identity, but does not satisfy the
derivation property \cite{flato2}.

In this section we wish to give an explicit
matrix realization of the
quantum Nambu bracket, which is skew-symmetric and obeys the
Fundamental Identity. Our simple example should be contrasted to
the example constructed in \cite{flato2}.

What do we mean by a quantum triple Nambu bracket? In general we want
an object $[F,G,H]$ which satisfies properties analogous
to the classical Nambu bracket $\{f,g,h \}$ as listed in
the previous section. (Here $f,g,h$ are
functions of three variables, and the nature of $F,G,H$ is left open
for the moment.) Thus $[F,G,H]$ is expected to
satisfy \cite{fi}, \cite{flato}, \cite{flato2}, \cite{filip}
\begin{enumerate}
\item Skew-symmetry
\EQ
[A_1, A_2,A_3]=(-1)^{\epsilon(p)} [ A_{p(1)}, A_{p(2)},A_{p(3)} ],
\label{skewsymmetry1}
\EN
where again $p(i)$ is the permutation of indices and $\epsilon(p)$
is the parity of the permutation,
\item Derivation
\EQ
[A_1A_2, A_3, A_4] =A_1[A_2, A_3,A_4] + [A_1, A_3,A_4]A_2 ,
\EN
\item Fundamental Identity (F.I.)
\EQA
[[A_1, A_2, A_3],A_4, A_5 ] &+& [A_3, [A_1, A_2,A_4],A_5]
\nonumber \\
+[A_3, A_4, [A_1, A_2, A_5]] &=&
[A_1,A_2,[A_3, A_4, A_5]].
\EQN
\end{enumerate}
(Note that the two-dimensional quantum Nambu bracket which satisfies
above properties is just the
usual commutator of matrices $[A,B] \equiv AB-BC$. In this case the
F.I. reduces to the Jacobi identity.)

First we point out that there exists a matrix realization of
the triple quantum Nambu bracket which satisfies the property of
skew-symmetry and the Fundamental Identity. To demonstrate this claim we
define a totally antisymmetric
triple bracket of three matrices A, B, C as
\EQ
[A,B,C] \equiv
({\rm tr} A)  [B,C] +(tr B)[C,A]+(tr C)[A,B] \label{31}.
\EN
Then ${\rm tr} [A,B,C]=0$, and if $C=1$, $[A,B,1]=N [A,B]$, where
$N$ is the rank of square matrices.

It is quite easy to prove the Fundamental Identity  now.
The F.I. is just an adjoint action involving matrices $X,Y,A,B,C$
$$
[X,Y,[A,B,C]]=[[X,Y,A],B,C]+[A,[X,Y,B],C]+[A,B,[X,Y,C]].
$$
Using the definition of the triple bracket, the L.H.S. is equal to
\EQA
&({\rm tr} A) [X,Y,[B,C]] +
c(A,B,C)&
= {\rm tr} A{\rm tr} X [Y,[B,C]]-{\rm tr} A {\rm tr} Y [X,[B,C]] 
\nonumber \\
&+c(A,B,C)&
\EQN
where $c(A,B,C)$ denotes the cyclic permutation of $A,B,C$.

On the other hand, the R.H.S. of the F.I. is equal to
\EQ
({\rm tr} X) [[Y,A],B,C]+({\rm tr} Y)[[A,X],B,C] +({\rm tr} A)[[X,Y],B,C]
+ c(A,B,C).
\EN
It is easy to see that
\EQ
({\rm tr} A) [[X,Y],B,C]+({\rm tr} B)[[X,Y,],C,A]+({\rm tr} C)[[X,Y],A,B]=0.
\EN
This is because each term is proportional to a product of two traces
of the three matrices $A,B,C$, and they cancel. For instance, from the
first term in the above equation we have $({\rm tr} A{\rm tr} B) [C,[X,Y]]$,
while from the second term we have $({\rm tr} B{\rm tr} A)[[X,Y],C]$ - the
two add up
to zero. Because of the above equation, the R.H.S. of the F.I. simplifies
to
$({\rm tr} X)[[Y,A],B,C]+({\rm tr} Y)[[A,X],B,C]+
c(A,B,C)$.

We can now see why the L.H.S. is equal to the R.H.S. From both
the L.H.S. and R.H.S.
 every term upon expansion is proportional to a product of two
traces, one comes from ${\rm tr} X$ or ${\rm tr} Y$, another from ${\rm
tr}
A, {\rm tr} B$ or ${\rm tr} C$. For example, on the L.H.S. there is a term
$$({\rm tr} A{\rm tr} X)[Y,[B,C]],$$
while from the R.H.S., we find
$$
({\rm tr} A {\rm tr} X)
( [B,[Y,C]]-[C,[Y,B]]).
$$
This term is equal to the one from the L.H.S. by virtue of the Jacobi  
identity.
The same is true for other terms, simply by doing permutations.
This completes the proof of the Fundamental Identity.

Given our example of a three-dimensional quantum Nambu bracket, let
us consider the following "gauge transformation"
\EQ
{\delta} A
\equiv i[X,Y,A], \label{32}
\EN
where we introduce the factor $i$, because we take all matrices to be
Hermitian. This transformation represents an obvious quantum form of the
three-dimensional volume preserving diffeomorphisms.

By the definition of
the triple bracket, the gauge transformation takes the following
explicit form
\EQ
\delta A=i\left( [({\rm tr} X)Y-({\rm tr} Y)X, A]+({\rm tr} A)[X,Y]\right)
\label{33}
\EN
where the first term in the parentheses is just the usual $su(N)$
gauge transformation, and the second term is apparently new.

The first important property of the gauge transformation, in addition
to satisfying the {\it generalized}
composition rule (or the F.I.), is that
$\delta {\rm tr} (AB)=0$ provided ${\rm tr} A={\rm tr} B=0$. To see this
look at
$$\delta {\rm tr} (AB) =i[{\rm tr} ([X,Y,A]B)+{\rm tr} ( [X,Y,B]A)]$$
which in turn is equal to
$$i[{\rm tr} ([({\rm tr} X)Y-({\rm tr} Y)X,A]B)+(A\leftrightarrow B)]=0.$$
The crucial point is that since ${\rm tr} A=0$, the last term in the gauge
transformation (\ref{33}) is absent, therefore the gauge transformation
is just an $su(N)$ gauge transformation.
Thus, we have the following  general
result:

If ${\rm tr} A_i=0, i=1, ... n$, then
\EQ
{\rm tr} (A_1A_2\dots A_n)
\EN
is gauge invariant.

Note that the gauge transformation of
a commutator does not satisfy the usual composition rule, namely
$$
[X,Y, [A,B]]\ne [[X,Y,A],B]+[A,[X,Y,B]] \label{34}.
$$
To see this, we only need to realize that the gauge transformation
of $[A,B]$ is just an $su(N)$ transformation, since ${\rm tr} [A,B]=0$.
On the other hand, there is an additional term on the R.H.S. of the above
inequality,
$({\rm tr} A)[[X,Y],B]+({\rm tr} B)[A,[X,Y]]$. If ${\rm tr} A={\rm tr} B  
=0$,
then the gauge transformation of the commutator $[A,B]$ satisfies the
composition rule.  Similar comments apply to $[[X^i,X^j],[X^l,X^k]]$,
since
although ${\rm tr} [X^i,X^j]=0$, $[X^i,X^j]$ itself does not satisfy
the composition rule.

We can also define a
triple quantum Nambu bracket involving fermionic matrices. If there is
only one
fermionic matrix involved in a triple bracket, the definition is the
same as in (\ref{31}), and the F.I. identity holds. If there are two
fermionic matrices, $\psi$ and $\lambda$, define
\EQ
[A,\psi,\lambda]=({\rm tr} A)\{\psi,\lambda\}+({\rm tr}\psi)[\lambda
,A]+({\rm tr} \lambda)[\psi, A],
\EN
where $\{\psi ,\lambda\}=\psi\lambda +\lambda\psi$, and
${\rm tr} \{\psi,\lambda\}=0$. Due to this property and the Jacobi identity
involving fermionic matrices, the F.I. still holds. Note that this triple
bracket is symmetric in $\psi$ and $\lambda$.

If all three matrices
are fermionic, we can define the following fermionic triple quantum Nambu
bracket
\EQ
[\psi,\lambda,\chi]=({\rm tr}\psi)\{\lambda,\chi\}
+({\rm tr}\lambda)\{\chi,\psi\} +({\rm tr}\chi)\{\psi,\lambda\}.
\EN
Obviously this triple bracket is totally symmetric in three fermionic
matrices.
Again we can repeat the same steps as before to show that the F.I.
holds.

Notice that the form of the gauge transformation (\ref{33})
indicates that a bosonic
Hermitian matrix $A$  can be transformed into a form proportional to
the unit $N \times N$ matrix as long as ${\rm tr} A\ne 0$. In other
words, since the gauge transformation is traceless,
one can show that a Hermitian matrix can be brought to the following form
$$A\rightarrow {1\over N}{\rm tr} A 1_N.$$
To prove this note that the first term in (\ref{33})
represents the usual $su(N)$ gauge transformation, and it
can be used to diagonalize $A$; the second term helps to balance all the
eigenvalues of $A$. Hence, to diagonalize $A$ first, choose $X$ to be a
Hermitian matrix
and $Y=1$, so that $\delta A=iN[A,X]$ - which is a standard $su(N)$ gauge
transformation. Once $A$ is diagonalized this way, choose $X,Y\in su(N)$,
so that the first term in (\ref{33})
is absent, and $\delta A=i ({\rm tr} A)[X,Y]$.
By choosing suitable $X$ and $Y$, we can let $[X,Y]$ be any element in
the Cartan algebra of $su(N)$, so that any residual of $A$ in this Cartan
algebra can be gauged away, thus proving the above statement.

One final remark: one might wonder whether there exists a ``c"-number
for our three-dimensional quantum  Nambu bracket (35), i.e. whether there
exists an
element
$E$ such
that
\EQ
[E, A, B] =0
\EN
for {\it arbitrary} elements $A, B$?
It appears that there is {\it no} solution for this
condition except $E=0$ for the triple Nambu
bracket defined above.

Notice also that if we consider the triple commutator of
the form
\[
[G, A, B]
\]
it follows from the property of skew-symmetry that this bracket is
invariant
under $ A \rightarrow A + G$ and $B \rightarrow B + G$. In other words,
$[G,G,B] = [G,A,G] =0$ precisely because of the property of skew-symmetry.


\section{\bf Cubic Matrices and the Nambu Bracket}


In this section we want to address the following question:
Is there a many-index matrix representation of the three-dimensional
quantum Nambu bracket? For example, the three-dimensional
classical Nambu bracket is naturally realized in terms of functions of
three variables.
Then it is natural to ask:
Is it possible to realize the three-dimensional
quantum Nambu bracket in terms of three-index objects $A_{ijk}$
(cubic matrices)?
It turns out that the answer to this question is positive.
In this section we give some explicit examples of
the three-dimensional quantum Nambu bracket written in terms of
three-index objects or cubic matrices.

Let us introduce the following generalization of the traces
\EQ
\< A \> \equiv \sum_{pm} A_{pmp},\qquad
\< A B \> \equiv \sum_{pqm} A_{pmq} B_{qmp},\qquad
\< ABC \> \equiv \sum_{pqrm} A_{pmq} B_{qmr} C_{rmp},
\label{e:trace}
\EN
which satisfy $\<AB\>=\<BA\>$ and $\<ABC\>=\<BCA\>=\<CAB\>$.
Let us furthermore define a triple-product
\begin{equation}
(ABC)_{ijk} \equiv \sum_{p} A_{ijp} \< B \> C_{pjk}
=\sum_{pqm} A_{ijp} B_{qmq} C_{pjk}. \label{t11}
\end{equation}
Given this triple-product we define the following skew-symmetric
quantum Nambu bracket
\begin{equation}
[A,B,C] \equiv
  (ABC) + (BCA) + (CAB)
- (CBA) - (ACB) - (BAC).
\label{e:NambuBracket}
\end{equation}
The middle index $j$ of $A_{ijk}$ can be treated
as an internal index for the matrix realization of the
three-dimensional quantum Nambu bracket we considered in
the previous section. Therefore we expect that the F.I. should be satisfied.
This indeed turns out to be the case.
Note also that
$\<(ABC)\> = \<B\>\<AC\> \neq \<ABC\>$ and
$\<(ABC)D\> = \<B\>\<ACD\>$.

Then by using the following relations
\EQA
((ABC)DE) &=& ((ADC)BE) = (AB(CDE)) = (AD(CBE)),\cr
(A(BCD)E) &=& (A(DCB)E), \label{t12}
\EQN
one can directly prove that
{\it the skew-symmetric Nambu bracket $(\ref{e:NambuBracket})$
with the triple-product $(\ref{t11})$ obeys the F.I. (34)}

The ``trace'' $\<AB\>$ has the property
$$\<\,[X,Y,A] B\,\> + \<\,A [X,Y,B]\,\> = 0,$$ provided $\<A\> = \<B\> = 0$.
Therefore, since $\<\,[A,B,C]\,\> = 0$ for any three-index 
objects $A$, $B$ and $C$,
{\it the trace of the product of Nambu brackets
$\<\,[A,B,C] [D,E,F]\,\>$ is gauge invariant.}
Notice that if we generalize the trace (\ref{e:trace}) as
\EQ
\< A^1 A^2 \cdots A^n \> \equiv \sum_{p_1,p_2,\cdots,m}
A^1_{p_1 m p_2} A^2_{p_2 m p_3} \cdots A^n_{p_n m p_1},\qquad n=1,2,\cdots,
\EN
we can also demonstrate that the trace of any product of Nambu brackets
$\<\,[A,B,C] [D,E,F] \cdots [X,Y,Z]\,\>$ is gauge invariant.

Let us also define
$I_{ijk} \equiv \delta_{ik}^{(j)}$,
where $\delta_{ik}^{(j)} =0$, if $i \neq k$, for any $j$ and
$\delta_{ik}^{(j)} =1$, if $i = k$, for any $j$.
Then
\EQA
(AIB) = \<I\> \sum_p A_{ijp} B_{pjk},&&
(IAB) = (BAI) = \<A\>B,
\cr
(IAI) = \<A\> \delta_{ik}^{(j)},\qquad &&
(IIA) = (AII) = \<I\> A,
\EQN
and
$[A,I,B] = \sum_p (A_{ijp} B_{pjk} - B_{ijp} A_{pjk})$.
Hence for any middle index $j$,
$[A,I,B]$ reduces to the usual commutator $[A^{(j)},B^{(j)}]$
for the matrices
$A^{(j)}_{ik} \equiv A_{ijk}$ and
$B^{(j)}_{ik} \equiv B_{ijk}$.

Finally we list other examples of triple-products $(ABC)_{ijk}$
which also satisfy the same relations as eq.\ (\ref{t12})
and hence lead to the F.I.
for the skew-symmetric Nambu bracket (\ref{e:NambuBracket})
\EQ
\sum_{pq} A_{ijp} B_{qjq} C_{pjk},\qquad
\sum_{pqmn} A_{ijp} B_{qmq} C_{pnk},\qquad
\sum_{pqmn} A_{inp} B_{qmq} C_{pjk}.
\EN


\section{Generalization: $n$-dimensional Quantum \\
Nambu Bracket}


We can now generalize and formalize our construction
for the $n$-dimensional quantum Nambu bracket
which obeys the $n$-dimensional Fundamental Identity $FI_n$:
\EQ
[X_1,\cdots,X_{n-1},[A_1,\cdots,A_n]]
= \sum_{i=1}^n[A_1,\cdots,[X_1,\cdots,X_{n-1},A_i],\cdots,A_n].
\EN

Let $A_i$ be operators and let $\<A_i\>$ be some $c$-numbers (``traces").
Let us define a generalized $n$-dimensional bracket
given an $(n-1)$-dimensional bracket as follows:
\EQ
[A_1,A_2,\cdots,A_n] \equiv
\sum_{i=1}^n (-1)^{i-1}
\<A_i\> [A_1,\cdots,\check A_i,\cdots,A_n] \label{311}.
\EN
Here $\check A_i$ stands for the term that is omitted.
Moreover, let us assume that
the ``trace" $\<A\>$ has the property
$\<\,[A_1,\cdots,A_n]\,\>=0$.
Note that
if the $(n-1)$-dimensional bracket is also related to the  
$(n-2)$-dimensional
bracket with the same "trace" $\<A\>$ as in the $n$-dimensional case,
then $[A_1,\cdots,A_n] = 0$.

By a straightforward computation,
one can show that
{\it
if an $(n-1)$-dimensional Nambu bracket
is skew-symmetric and satisfies $FI_{n-1}$
then the corresponding $n$-dimensional bracket
is also skew-symmetric and obeys $FI_n$} (52).


\section{Conclusions}


In this article we have constructed several explicit examples of
the quantum Nambu bracket in terms of square matrices and
three-index objects - cubic matrices.
Our examples satisfy two crucial properties of the classical Nambu bracket:
skew-symmetry and the Fundamental Identity.
Our results should be compared to the existing literature on the
deformation quantization approach to Nambu mechanics \cite{flato2}.
We have also discussed the generalization of our approach to the
$n$-dimensional quantum Nambu bracket.
Notice that it still remains on open question whether
all three properties of the classical Nambu bracket -
skew-symmetry, derivation and the F.I. -
can be realized
in terms of square or cubic matrices.

We wish to conclude this article by outlining a few
possible applications of our formulation.
One possible application of our results is in relation to
the yet unknown mathematical structure of
the space-time uncertainty relation in M-theory \cite{stu}.
There seems to exists a naive similarity between the form of the
space-time uncertainty relation in M-theory and the three-dimensional
quantum Nambu bracket \cite{stu}.
It would be interesting to formulate this intuitive
relation more precisely. Also, in view of the fact that space-time
uncertainty principle captures important  qualitative features of Matrix
theory \cite{matrix} it would be very interesting to see  whether our
results could
be relevant for the still unsolved problem of covariantization of Matrix
theory \cite{hadm}. One universal technical aspect of any attempt
to covariantize Matrix theory is to promote all spacetime coordinates
to matrices, and then to gauge fix a matrix, say the light-cone time
matrix, to proportional to the identity matrix. This demands enlarging
the gauge symmetry. Our model of Nambu bracket presented in sect. 3
gives rise to gauge symmetry large enough to achieve this goal.

Furthermore, given  the fact that the light-cone action for a super
p-brane is
invariant under the volume preserving diffeomorphisms \cite{volume}
and given the
obvious
relationship between p-dimensional ($p \ge 2$) volume preserving
diffeomorphisms and p-dimensional classical Nambu brackets
\cite{membrane}, \cite{area}, \cite{volume}, we expect
that our results
should shed some light on the quantization problem of p-branes in
M-theory and the unknown structure of the non-Abelian antisymmetric
tensor gauge theory that emerges in the case when $p \ge 3$ \cite{volume}.

We hope to return to some of these problems in the future.


\vspace{1cm}
\noindent
{\bf Acknowledgements}


It is our pleasure to thank T. Banks, I. Bars, O. Bergman,
S. Chaudhuri, E. Gimon, M. G\"{u}naydin,  P. Horava,
T. H\"{u}bsch,  J. Minahan, Y. Nambu, J. Polchinski, K. Pilch,
J. Schwarz, L. Smolin, P. Wiegmann, E. Witten and C. Zachos for
interesting comments
and
discussions. D. M. would like to thank Tokyo University and
the Yukawa Institute for Theoretical Physics for their hospitality
during the final stage of this work.
The work of M.L. is supported in part by DOE
grant DE-FG02-90ER-40560 and NSF grant PHY 91-23780.
The work of D. M. is supported in part by DOE grant
DE-FG03-84ER40168 and by a National Science Foundation grant
NSF9724831 for collaborative research between USC and Japan.
The work of
T.Y. is supported in part
by Grant-in-Aid for Scientific Research (No. 09640337) and
Grant-in-Aid for International Scientific Research (Joint
Research, N0. 10044061) from the Ministry of Education, Science and
Culture.

\vspace{1cm}

\noindent
{\large {\bf Appendix}}
\renewcommand{\theequation}{A.\arabic{equation}}
\appendix
\setcounter{equation}{0}
\vspace{0.3cm}

\noindent
Sufficient Condition for the F.I.

\vspace{0.3cm}

In this appendix
we will give a sufficient condition for the F.I. (34).
To find this, it is worth to recall the classical case:
Why does the Nambu Poisson bracket obey the classical F.I.?

For the usual Poisson bracket
$\{A,B\}=A_pB_q-B_pA_q$ with $A_p\equiv\partial_p A$,
it is natural to define a non-commutative
product as $(AB)\equiv A_pB_q$.
Then
\begin{eqnarray}
((AB)C) &=& A_pB_{pq}C_q + A_{pp}B_qC_q,\cr
(A(BC)) &=& A_pB_pC_{qq} + A_pB_{pq}C_q.
\end{eqnarray}
We thus distinguish three kinds of triple-products:
$A_pB_{pq}C_q$, $A_{pp}B_qC_q$ and $A_pB_pC_{qq}$.
The second triple-product is symmetric under the interchange of $B_q$ and
$C_q$; the third is symmetric under the interchange of $A_p$ and
$B_p$.

We can generalize this to the quantum case.
Let us define three kinds of triple-products
$(ABC)^M$  $M=0,23,12$ which possess
symmetries of their classical
counterparts
$(ABC)^{23}=(ACB)^{23}$ and
$(ABC)^{12}=(BAC)^{12}$.
Then a sufficient condition for the Jacobi identity is
\begin{eqnarray}
((AB)C) &=& (ABC)^0 + (ABC)^{23},\cr
(A(BC)) &=& (ABC)^{12} + (ABC)^0.
\end{eqnarray}
For the usual square matrices,
only one  triple-product, $(ABC)^0$, is non-zero and
hence the sufficient condition is nothing but the property of associativity
$((AB)C)=(A(BC))$, as expected.

In the case of a three-dimensional Nambu bracket,
the associativity is too weak for the F.I.,
and the corresponding sufficient condition turns out to be more complicated.
If we define the classical Nambu triple-product as
$(ABC)\equiv A_pB_qC_r$, in analogy with the treatment of the
Poisson bracket, then
\begin{eqnarray}
((ABC)DE) &=& A_pB_qC_{pr}D_qE_r + A_pB_{pq}C_rD_qE_r +  
A_{pp}B_qC_rD_qE_r,\cr
(A(BCD)E) &=& A_pB_{pq}C_qD_rE_r + A_pB_pC_{qq}D_rE_r +  
A_pB_pC_qD_{qr}E_r,\cr
(AB(CDE)) &=& A_pB_qC_pD_qE_{rr} + A_pB_qC_pD_{qr}E_r + A_pB_qC_{pr}D_qE_r.
\end{eqnarray}
Thus we have to consider six kinds of quintuple-products which possess
various obvious symmetries.

Again we can generalize this to the quantum case.
Let us consider six kinds of quintuple-product with symmetries
analogous to their classical counterparts
\begin{eqnarray}
(ABCDE)^{24} &=& (ADCBE)^{24}, \cr
(ABCDE)^{35} &=& (ABEDC)^{35}, \cr
(ABCDE)^{13} &=& (CBADE)^{13}, \cr
(ABCDE)^{24,35} &=& (ADCBE)^{24,35} = (ABEDC)^{24,35},\cr
(ABCDE)^{12,45} &=& (BACDE)^{12,45} = (ABCED)^{12,45},\cr
(ABCDE)^{13,24} &=& (CBADE)^{13,24} = (ADCBE)^{13,24}.
\end{eqnarray}
Thus we have found that the sufficient condition for the F.I.
for the skew-symmetric Nambu bracket (\ref{e:NambuBracket})
is
\begin{eqnarray}
((ABC)DE) &=& (ABCDE)^{24} + (ABCDE)^{35} + (ABCDE)^{24,35},\cr
(A(BCD)E) &=& (ABDCE)^{35} + (ABCDE)^{12,45} + (ACBDE)^{13},\cr
(AB(CDE)) &=& (ABCDE)^{13,24} + (ABCDE)^{13} + (ABCDE)^{24}.
\end{eqnarray}

Note  that the condition of associativity
$((ABC)DE) = (A(BCD)E) = (AB(CDE))$
does not guarantee the F.I. unlike the case of matrices, where the Jacobi
identity is implied by the associativity of the matrix product.
The examples in section 4.\ are such that
only one  quintuple-product $(ABCDE)^{24}$ is non-zero and
$(A(BCD)E)$ is a trivial quintuple-product
which implies  $(A[BCD]E)=0$.


\end{document}